\documentclass[prb,showpacs,twocolumn,floatfix]{revtex4}
\usepackage{graphicx}
\begin{document}
\def\rhov{{\mbox{\boldmath{$\rho$}}}}
\def\tauv{{\mbox{\boldmath{$\tau$}}}}
\def\Lambdav{{\mbox{\boldmath{$\Lambda$}}}}
\def\sigmav{{\mbox{\boldmath{$\sigma$}}}}
\def\xiv{{\mbox{\boldmath{$\xi$}}}}
\def\chiv{{\mbox{\boldmath{$\chi$}}}}
\def\rhov{{\mbox{\boldmath{$\rho$}}}}
\def\phiv{{\mbox{\boldmath{$\phi$}}}}
\def\piv{{\mbox{\boldmath{$\pi$}}}}
\def\psiv{{\mbox{\boldmath{$\psi$}}}}
\def\oh{{\scriptsize 1 \over \scriptsize 2}}
\def\ot{{\scriptsize 1 \over \scriptsize 3}}
\def\of{{\scriptsize 1 \over \scriptsize 4}}
\def\tf{{\scriptsize 3 \over \scriptsize 4}}
\title{Charge and Spin Ordering in the Mixed Valence Compound
LuFe$_2$O$_4$}
\author{A. B. Harris$^1$ and T. Yildirim$^2$}

\affiliation{[1] Department of Physics and Astronomy,
University of Pennsylvania, Philadelphia, PA 19104}
\affiliation{[2] NIST Center for Neutron Research, National Institute of
Standards and Technology, Gaithersburg, MD 20899 and
Department of Materials Science and Engineering,
University of Pennsylvania, Philadelphia, PA 19104}
\date{\today}

\begin{abstract}
Landau theory and symmetry considerations lead to a unified
treatment of charge and spin ordering in the mixed valence compound
LuFe$_2$O$_4$.  The unusual evolution of charge ordering
is attributed to interactions between charges
and phonons whose frequencies are calculated from first principles.
\end{abstract}
\pacs{75.25.+z,75.10.Jm,75.40.Gb}
\maketitle

In the past few years there has been an explosion of interest in
systems which display simultaneous magnetic and ferroelectric
ordering.[\onlinecite{TKIM,LCC,LAWES,MK}] In many of these compounds two
magnetic phase transitions are required to produce the necessary
lowering of symmetry to allow ferroelectricity[\onlinecite{ABH}].
A slightly different scenario is presented by members of the mixed valence
family RFe$_2$O$_4$, where R is a rare earth, in which half the Fe ions
carry charge +2e and half carry charge +3e and here we consider
the case R =  Lu.  LuFe$_2$O$_4$ (LFO) has a trigonal $R\overline 3 m$
crystal structure[\onlinecite{ACTA}]
and the Fe ions form
triangular lattice layers (TLL's) arranged in bilayers with two Fe ions
per rhombohedral primitive unit cell, as shown in Fig. 1a.  At 
temperatures above 500K, the valence electrons can thermally hop so that
all sites appear to have charge 2.5e.[\onlinecite{TANAKA}]
As the temperature is reduced from 500K, charge ordering (CO) correlations
develop at wave vectors associated with ``root 3" ordering within each
TLL (see Fig. 1b) and correspond to CO within each bilayer such that
the dipole moment of all bilayers are parallel.[\onlinecite{ANGST}]
(We call these ``ferro" (F) configurations.) At 
$T=T_{\rm CO} =350$K[\onlinecite{IKEDA,FNT}] there is a continuous transition
at which CO develops 3 dimensional (D) long range order without
any long range magnetic order.[\onlinecite{ANGST}] However, it is striking that
the long-range CO occurs in an ``antiferro" (AF) 
configuration in which the dipole
moments of adjacent bilayers are oppositely
oriented.[\onlinecite{ANGST}] Angst {\it et al.}
argue that the previous report of a spontaneous polarization[\onlinecite{IKEDA}]
in this CO phase may be an artefact of the small applied electric field.
As the temperature is further reduced, the spontaneous polarization
${\bf P}$ (in a small electric field)
quickly saturates and is constant between
320K and 240K[\onlinecite{IKEDA}].  At $T_c=240$K there is a transition
in which a ferrimagnetic state appears which, at low temperature, has a
moment per unit cell which is very nearly 1/3 of the moment that would occur
if all spins were parallel[\onlinecite{IIDA,CHRIST}].  At this transition
${\bf P}$ abruptly begins to further increase so that[\onlinecite{IKEDA}]
\begin{eqnarray}
P(T)=P(T_c)+a (T_c-T)^{\beta'}\ , \hspace{0.5 in} \beta' \approx 1 \ ,
\label{PEQ} \end{eqnarray}
where $\beta'$ appears to be somewhat smaller than 1.
At $T=175$K there is a transition which 
may be of magnetic origin.[\onlinecite{CHRIST}] The major
questions raised by these behaviors are
(a) what does Landau theory say about the incommensurate
order wave vector?
(b) why are the dominant fluctuations for $T> T_{\rm CO}$ F whereas
for $T<T_{\rm CO}$ the ordered phase is AF?
and (c) what form does the coupling between magnetic 
order and the polarization take and what restrictions does
symmetry place on the direction of the polarization of a bilayer?

\begin{figure}[ht]
\begin{center}
\includegraphics[width=7.5 cm]{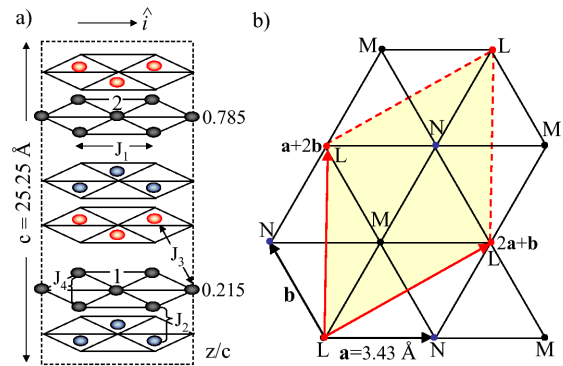}
\caption{\label{TET} (Color online) a) Fe ions in LFO with charge-charge
interactions $J_n$ indicated.  The hexagonal (conventional) unit cell contains
three bilayers configured so layers are stacked in the order ABCABC (A=red,
B=black, C=blue) with two sites in the rhombohedral unit cell labeled "1" 
and "2" which are related by a center of inversion symmetry.  b) root 3 
structure of charge or spin ordering in a single TLL, where
$L=\sigma \cos \psi$, $M=\sigma \cos (\psi + 2 \pi/3)$, and
$N= \sigma \cos(\psi-2 \pi /3)$, where $\sigma$ is the amplitude
and $\psi$ the phase of the 
order parameter. The shaded rhombus is the broken symmetry unit cell.}

\end{center}
\end{figure}

We first consider question (a).  A major advance in understanding CO
was due to Yamada {\it et al}[\onlinecite{YAMADA}]
who presented  calculations which could explain the unique zig-zag
X-ray pattern of spots at $T=270$K[\onlinecite{YAMADA}] at
vectors (in hexagonal rlu's)[\onlinecite{FN1}]
\begin{eqnarray}
{\bf q}_1 &=& (h + 1/3 + \delta'/3, h+1/3+\delta'/3,
\tau ) \ , \nonumber \\
{\bf q}_2 &=& (h - 2/3 - 2 \delta'/3, h+1/3+\delta'/3,
\tau  - 2) \ , \nonumber \\
{\bf q}_3 &=& (h + 1/3 + \delta'/3, h-2/3-2\delta'/3,
\tau - 1) \ ,
\label{EQB} \end{eqnarray}
where $\tau=3(n+1/2)$.  Here $h$ and $n$ are integers and $\delta' = 0.0081$.
However, surprisingly, Yamada {\it et al.} claimed that 3D order would 
not occur even in the presence of the interactions
$J_2$ and $J_3$ between adjacent bilayers.

We start with a Landau analysis of CO using the
lattice gas model [\onlinecite{YAMADA}] in which
one introduces a variable
at the $n$th site of the rhombohedral unit cell at ${\bf R}$, $Q_n({\bf R})$,
which assumes the value $+1$ ($-1$) if the site is occupied
by a Fe$^{3+}$ (Fe$^{+2}$) ion. As shown in Fig. 1, one has interactions
$J_1$ and $J_4$ within each TLL, and interactions
$J_2$ and $J_3$ between nearest neighbors on adjacent bilayers.
$J_1$ is AF (positive).  As argued in Ref. \onlinecite{YAMADA}, rather than
use a long-ranged Coulomb interaction[\onlinecite{NAGPRL}],
we invoke a short-ranged interaction 
strongly screening by the large dielectric constant[\onlinecite{EPS}]
of the bulk.  The Fourier transformed variables, $Q_n({\bf q})$,
satisfy $Q_n({\bf R})= \sum_{\bf q} Q_n({\bf q})
\exp (-i {\bf q} \cdot {\bf R})$.
At quadratic order the Landau free energy is
\begin{eqnarray}
F &=& \frac{1}{2} \sum_{\bf q} \sum_{n,m=1}^2
F_{nm}({\bf q}) S_n({\bf q})^* S_m({\bf q}) \ ,
\label{EQA} \end{eqnarray}
where $S_n({\bf q}) = \langle Q_n({\bf q}) \rangle$, where
here and below
$\langle ... \rangle$ indicates a thermal average, and
$F_{nm}({\bf q}) =F_{mn}({\bf q})^*$.  Under spatial inversion, 
${\cal I}$, ${\cal I}S_1({\bf q}) = S_2({\bf q})^*$, which
implies that $F_{11}=F_{22}$.  In a simple approximation
\begin{eqnarray}
F_{nm}({\bf q}) = cT \delta_{nm} + \sum_{\bf R} J(0,m;{\bf R},n)
\exp(i {\bf q} \cdot {\bf R}) \ ,
\end{eqnarray}
where we set $k_B=1$, $J(0,m;{\bf R},n)$ is the interaction between
sites $m$ in the rhombohedral unit cell at the origin and $n$ in the
rhombohedral unit cell at ${\bf R}$,
and $c$ is a constant of order unity.  In Cartesian coordinates
\begin{eqnarray}
F_{11}&=& cT + 2J_1 [ \cos(aq_x) + 2 \cos(aq_x/2) \cos(\sqrt 3 aq_y/2)] 
\nonumber \\ &+ & 2J_4[ \cos(\sqrt 3 a q_y) + 2 \cos (3 a q_x/2)
\cos( \sqrt 3 a q_y/2)] \ ,
\nonumber \\
F_{21}&=& J_2 e^{-2icq_z/3} \Lambda(q_x,q_y)^*
+ J_3 e^{-icq_z/3} \Lambda(q_x,q_y) \ ,
\end{eqnarray}
where 
\begin{eqnarray}
\Lambda(q_x,q_y) &=& 2 e^{iq_y a \sqrt 3 /6} \cos(aq_x/2) +
e^{-iaq_y \sqrt 3 /3} \ .
\end{eqnarray}
To identify the critical wave vector(s) that minimizes
the eigenvalue of ${\bf F}$ we
set $aq_x = 4 \pi /3 + \eta$ and $aq_y=\rho$.  The wave vector
for $\eta = \rho =0$ (which we refer to as the $X$ point) gives the
root 3 structure (see Fig. 1b) in each TLL and it is expected that
the critical wave vector will have small $\eta$ and $\rho$. 
To lowest sensible order, we find that
\begin{eqnarray}
F_{11} &=& [ -3 + (3/4) \delta^2 - (\sqrt 3/8) \delta^3
\cos (3 \phi)] J_1 \nonumber \\ &+&
[ 6 - (9/2) \delta^2]J_4 + {\cal O} (\delta^4)  \ ,
\end{eqnarray}
where $\eta= \delta \cos \phi$ and $\rho = \delta \sin \phi$. Also
\begin{eqnarray}
\Lambda = - (\sqrt 3 /2) [\eta + i \rho] \equiv (- \sqrt 3 /2) 
\delta e^{i \phi} \ .
\end{eqnarray}
This is an important result which shows that generically, the $X$ point
(which can be stable for a 2D triangular lattice) is not stable in the
present case due to the off-diagonal terms in $\Lambda$ which are linear
in the wave vector.  This is an example of the Lifshitz condition which
states that one can only have a temperature independent commensurate wave 
vector in a continuous transition if the wave vector has sufficiently high
symmetry.[\onlinecite{LL}]

Thus the critical eigenvalue, $\lambda({\bf q})$ is
\begin{eqnarray}
&& \lambda({\bf q}) = F_{11} \nonumber \\
&& - \frac{\sqrt 3 \delta }{2}
\Bigl[ J_2^2 + J_3^2 + 2 J_2J_3 \cos (2 \phi + cq_z/3) \Bigr]^{1/2} \ .
\label{EQ6} \end{eqnarray}
We regard $\lambda({\bf q})$ as a function of $\phi$, $\delta$ and
$\tilde q_z \equiv 2 \phi + cq_z/3$.  Minimizing with respect to $\phi$ 
yields $\phi= \phi_n = 2 n \pi /3$, for $n=0,1,2$.
For $\lambda ({\bf q})$ to be minimal, the square bracket in
Eq. (\ref{EQ6}) should be maximal.  For $J_2J_3$ negative, the minima
occur for $\cos(2 \phi_n+cq_z/3)= -1$ in which case one finds the wave
vectors of Eq. (\ref{EQB}) as was found in 
Ref. \onlinecite{YAMADA}.[\onlinecite{FN1}]
For $J_2J_3 > 0$, the minima occur for $\cos(2 \phi_n+ ck_z/3)=+1$,
in which case one finds the wave vectors given
by Eq. (\ref{EQB}) but with $\tau=3n$.
Since fluctuation corrections to mean field theory will
be smaller at higher temperature, we use the
scattering at high temperatures (dominantly $\tau$ is integral)
to deduce that $J_2J_3$ is positive and leave the
CO phase data ($\tau$ is half-integral) to be explained
as a fluctuation effect.

Minimizing $\lambda ({\bf q})$ for $J_2J_3$ positive yields
\begin{eqnarray}
|J_2+J_3|/J_1= \sqrt 3 \delta=4 \pi \delta'/\sqrt 3 = 0.064 \ .
\label{J23EQ} \end{eqnarray}
Also for $F_{11}$ to be minimal at $\delta=0$ requires that
$J_1-6J_4$ be positive.  These relations indicate that the
interaction $J(r)$ decays with distance much more rapidly than
do bare Coulomb interactions. So we neglect $J_4$ in comparison
to $J_1$.  We set $\lambda({\bf q})=0$ to obtain the mean-field
value $T_{\rm CO} \approx 3J_1$. Since quasi-2D fluctuations can
be severe we adopt the estimate $J_1=40$ meV = 500 K.

We now discuss question (b) concerning
the competition between F fluctuations and AF ordering.
Note that $\lambda ({\bf q})$ depends on $q_z$ only through the
small interactions $J_2$ and $J_3$ with $0< J_3 \ll  J_2$. 
(The separation for $J_3$ ($J_2$) is $6.30\AA$ ($3.15\AA$) which
leads us to posit that $J_3/J_2$ is of order 0.1.) Thus
Eq. (\ref{J23EQ}) indicates that $J_2/J_1=0.06$ and
hence that $J_3=0.1J_2=0.3$ meV.
The difference in $T_{\rm CO}$ for F and AF CO, $\Delta T$,
is, from Eq. (\ref{EQ6}),
\begin{eqnarray}
\Delta T &=& T_{\rm CO}(q_z=0) - T_{\rm CO}(q_z=3 \pi /c)
= \sqrt 3 \delta J_3 \ ,
\label{DTEQ} \end{eqnarray}
from which we estimate that $\Delta T = 0.06J_3 = 0.25$ K.  Accordingly,
it makes sense to consider a theory in which F CO at $q_z=0$, with
order parameter (OP) $\sigma_B$, competes with AF CO at 
$q_z=3 \pi /c$, with OP $\sigma_A$.  Our conclusion that
$\Delta T$ is small is consistent with the suggestion of Angst
{\it et al.} that a spontaneous polarization is
observed[\onlinecite{IKEDA}] because of the presence of a small
applied electric field is enough to favor F stacking of bilayers
over AF stacking.  We point out that to obtain a polarization 
the small electric field has to also favor a commensurate state
with $\delta=0$ in preference to an incommensurate state.
We thus consider a free energy of the form[\onlinecite{BRUCE}]
\begin{eqnarray}
F &=& \frac{1}{2} r_A\sigma_A^2 + \frac{1}{2} r_B \sigma_B^2
+ u [\sigma_A^2 + \sigma_B^2]^2 + v \sigma_A^2 \sigma_B^2 \ .
\label{EQBB} \end{eqnarray}

\begin{figure}[ht]
\begin{center}
\includegraphics[width=8.0 cm]{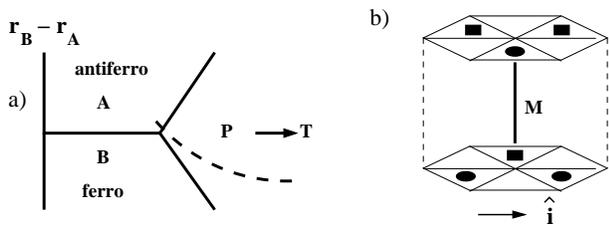}
\caption{\label{BI} a) Mean-field phase 
diagram[\onlinecite{BRUCE}] near the bicritical point of
Eq. (\ref{EQBB}) for $v>0$. Since the dashed trajectory in the
disordered (P) phase is closer to the F phase than to the
AF phase, F fluctuations dominate AF fluctuations
in the P phase. b) A single CO bilayer with a mirror plane {\bf M}
perpendicular to the wave vector which is along
$\hat i$.  Squares are Fe$^{3+}$'s and circles are Fe$^{2+}$'s.}
\end{center}
\end{figure}

We propose the trajectory shown in Fig. 2a
to explain dominant F fluctuations in the disordered phase but 
with condensation into an AF ordered state.  This trajectory
implies a temperature dependent renormalization of
the quadratic coefficients in Eq. (\ref{EQBB}) which favors
AF fluctuations over F fluctuations.  Quartic
terms for a single OP arise from implementation of the
fixed length constraint and are the same for F ($\sigma_B$) and
AF ($\sigma_A$)
fluctuations.  So these terms lead to a renormalization of $T_{\rm CO}$,
but do not explain the {\it slope} of the trajectory in Fig. 2a.
Accordingly we analyze a free energy of the form 
\begin{eqnarray}
V &=& a |\sigma^2| Y + (1/2) \chi_Y^{-1} Y^2 \ ,
\label{VEQ} \end{eqnarray}
where $Y$ is a noncritical variable 
and $\sigma$ is either the OP $\sigma_A$ or $\sigma_B$.
Minimizing with respect to $Y$ we obtain a renormalized
quartic interaction
\begin{eqnarray}
V &=& - (1/2) \chi_Y a^2 |\sigma^2|^2\ .
\label{FERROEQ} \end{eqnarray}
To leading order in the fluctuations we replace $|\sigma^2|^2$ by
$4 \langle |\sigma^2| \rangle |\sigma^2|$ and $V$ leads to a correction
to $r$ of
\begin{eqnarray}
\delta r &=& - 4 \chi_Y a^2 \langle |\sigma^2| \rangle \ .
\label{DREQ} \end{eqnarray}
which is temperature dependent, increasing in magnitude as the 
temperature is lowered.  It is crucial to show that
such terms favor AF fluctuations and thus
lead to a trajectory with the slope shown in Fig. 2a.

\begin{figure}[ht]
\begin{center}
\includegraphics[width=7.0 cm]{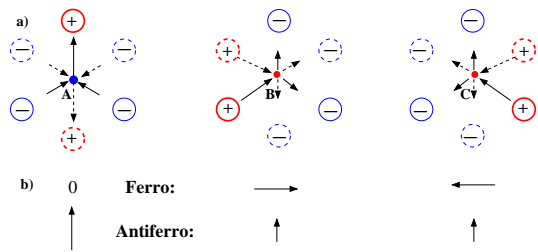}
\caption{(Color online) \label{CHOR} a) The in-plane components of
forces on sites L, M, and N in TLL$_0$ (as in Fig. 1) in the F configuration.
The solid line vector forces are from the charges (solid circles) in the
TLL above TLL$_0$ and  the dashed line vector forces are from the charges
(dashed circles) in the TLL below TLL$_0$.  The charges at A, B, and C
are negative, positive, and positive, respectively. The larger
dots and heavier lined circles are charges of twice the magnitude
of the smaller dots and lighter lined circles. 
b) The net force on the sites assuming the separations between all
planes are the same.   For the AF configuration
the dashed forces are reversed and the resulting total forces are listed
as ``Antiferro:".}
\end{center}
\end{figure}

Accordingly, we treat charge-phonon interactions where the phonon
amplitude $u$ is the noncritical variable and from Eq. (\ref{VEQ})
we see that $u$ must be an even symmetry phonon.
To see that this interaction
favors the AF configuration look at Fig. 3
where we show the net force on a given TLL (called TLL$_0$) from the
neighboring TLL's. One see that the forces from the neighboring TLL's
tend to cancel (add up) if the adjacent TLL's are in an F (AF)
configuration.  Table I shows the results of a first-principles
[\onlinecite{TY}] calculation which gave optical phonons at
11 meV and, for the transverse $g$ type needed for this mechanism,
at 20 meV.

\begin{table}
\caption{Calculated zone center phonon energies (in meV) and their
symmetry labels.  The $\Gamma_u$ and $\Gamma_g$ modes are
odd and even under inversion,
respectively. The nondegenerate (A) and doubly degenerate (E) modes
correspond to motion along the $c$-axis and within the $ab$-plane,
respectively.}

\vspace{0.1 in}

\begin{tabular} {|c| c c c c c c |} \hline
$\Gamma$  & $E_u$ & $E_g$ & $A_{2u}$ & $A_{1g}$ & $A_{2u}$ & $E_u$ \\
\ \ Energy \ \  & 11.41 & 19.99 & 20.02 & 31.48 & 38.46 & 41.20 \\  
\hline
$\Gamma$ & $A_{1g}$ & $E_g$ & $A_{2u}$ & $E_u$ & $E_g$ & $A_{1g}$ \\ 
Energy & 53.25 & 54.73 & 57.69 & 58.79 & 62.62 & 84.96 \\
\hline
\end{tabular}
\end{table}

To estimate the effect of this coupling we write
\begin{eqnarray}
F_{\rm Ph} &=& \oh \sum_i M \omega_D^2 {\bf u}_i^2 +
\sum_{ij} {\bf u}_{ji} \cdot \nabla_{\bf r} J({\bf r}_{ij}) 
Q_i Q_j \ ,
\end{eqnarray}
where $J({\bf r}_{ij})$ is the heavily screened Coulomb
interaction.  As in Fig. 1, the $Q$'s are given in terms
of the order parameter $\sigma_X$, where $X$
indicates either the F or AF configuration of TLL's.
Because the transverse motion of planes is relatively 
soft, we consider displacements to lie within the TLL$_0$ plane. 
When minimized with respect to $u_i$, the free energy is
\begin{eqnarray}
F_{\rm Ph} &=& - \frac{z^2 |\sigma_X|^4} {2 M \omega_D^2}
\left( \frac{J}{r} \right)^2 \left( \frac{r_\parallel}{r} \right)^2
\xi^2 \ ,
\end{eqnarray}
where $\xi = (r/J) (dJ/dr)$, $r_\parallel$ is the component of
$r$ within the TLL, and $z$ is the effective number of nearest
neighbors.  The actual number of nearest neighbors is 6, but since
the forces do not all add up, we take $z=3$ for the F configuration
and $z=0$ for the AF configuration where the forces from
adjacent TLL's nearly cancel.
We set $\hbar \omega_D = 20$ meV, $r=5\AA$, $r_\parallel=2\AA$, and
use our previous estimate that $J= 1 $ meV.  The factor $\xi$ depends
on how rapidly the interaction decreases with distance.  For
bare Coulomb interactions $\xi=-1$.  But we are far from that regime.
We take $\xi=-10$, which is a value often found in insulators.\cite{BLOCH}
Also the Fe mass is $M=60$ amu.  Thus
\begin{eqnarray}
F_{\rm Ph} &=& - F_0 |\sigma_{\rm AF}^4| \ ,
\end{eqnarray}
where $F_0 \approx 0.1$meV.  Equations (\ref{FERROEQ}) and (\ref{DREQ}),
yield 
\begin{eqnarray}
\delta r_{\rm AF} \approx - 1 {\rm meV} \langle |\sigma_{\rm AF}^2| \rangle \ ,
\end{eqnarray}
which is enough to shift ordering from F to AF below $T_c$.

We now analyze magnetic ordering using a Landau free energy assuming
only the $z$-components of spin are relevant.  Symmetry dictates that
the free energy quadratic in the magnetic variables, $S_n({\bf q})$,
has 
exactly
the same form as Eq. (\ref{EQA}).
The important difference from CO is that for spin ordering one
has significant reflections only at integer values
of $c k_z/(2 \pi)$.  Therefore in Eq. (\ref{EQ6}) it is clear that
$J_2J_3$ is positive, so that the minima are given
by Eq. (\ref{EQB}) with $\tau=3n$.  Although no such
zig-zag pattern has been reported,
such a phenomenon may be hard to detect if $\delta'$ is small,
as it was overlooked by many studies of CO. We note that this spin
incommensurability need not be coupled to that of CO.

Now we consider the situation when
the incommensuration $\delta$ is zero.
At low temperature CO is such that two thirds of the sites
in a TLL unit cell are Fe$^{2+}$ and one third Fe$^{3+}$, or vice versa.
For spin ordering two thirds of the spins should be fully up and one third
fully down.  Note that the commensurate state
with a {\it single} Fourier component is
not satisfactory because it does not permit the expected nonzero
total charge or magnetic moment in each TLL.  To get that
we invoke the interaction
\begin{eqnarray}
V &=& - a \sigma({\bf q})^3  \sigma(0,0,3q_z) + (1/2) \chi^{-1}
\sigma(0,0,3q_z)^2 \ ,
\label{SECOND} \end{eqnarray}
where $a$ is a constant.
This term is allowed because $3 {\bf q} + 3 q_z \hat k$ is a
reciprocal lattice vector.
Here we assume that $q_z$ for the order parameter $\sigma$
is either F or AF, so that $3cq_z/(2 \pi)$ is
either an integer or a half-integer. Minimization of $V$
gives $\sigma(0,0,3q_z)=a \chi \sigma({\bf q})^3$. 
It is also worth noting that the phase $\psi$ of the OP
(see Fig. 1b), although not fixed by scattering data, is fixed
by a sixth order term in the free energy,
\begin{eqnarray}
V_6 &=& C[\sigmav({\bf q})^6 + {\sigmav({\bf q})^*}^6]
= 2C |\sigma|^6 \cos (6 \psi) \ ,
\label{V6EQ} \end{eqnarray}
where $C$ is a constant and $\sigmav = |\sigma| \exp(i \psi)$.
For $C<0$, $V_6$ is minimal for
$\psi=n \pi /3$, where $n$ is an integer.

We now discuss question (c), starting with the experimental result
of Eq. (1) for $P(T)$ measured in a small electric field.[\onlinecite{IKEDA}]
Since $P(T)$ is already ordered, the magnetoelectric free energy is of the form
\begin{eqnarray}
F &=& (1/2) [\chi_\parallel^{-1} \delta P_\parallel^2
+ \chi_\perp^{-1} \delta P_\perp^2] + a {\sigma_S}^2 
[{\bf P}(\sigma_S)]^2\ ,
\end{eqnarray}
where ${\bf P}_\parallel={\bf P}(\sigma_S=0)$,
$\sigma_S$ is the ferrimagnetic OP, $\chi$ is the dielectric
susceptibility, $a$ is a constant, and
$\delta {\bf P} \equiv {\bf P}(\sigma_S) - {\bf P}(\sigma_S=0)$.
This free energy is invariant
under time reversal ($\sigma_S \rightarrow -\sigma_S$)
and inversion (${\bf P} \rightarrow -{\bf P}$). Minimizing with
respect to $\delta {\bf P}$ we find that $\delta P_\perp=0$ and
\begin{eqnarray}
P_\parallel(\sigma_S) &=& P(\sigma_S=0) + 2 a \chi_\parallel 
P(\sigma_S=0) \sigma_S^2 \ ,
\end{eqnarray}
so that if $\sigma_S \sim (T-T_c)^\beta$, then 
$\delta P_\parallel \sim (T-T_c)^{2 \beta}$.  This result comports
quite nicely with experiment, since this exponent at $T_c$ 
appears to be close to 1.[\onlinecite{IKEDA}]

Finally, we comment on the direction of ${\bf P}$ for
a single bilayer.  As mentioned, the polarization is zero
unless $\delta=0$ (the system is commensurate).
In Fig. 2b we show a bilayer with commensurate root 3 CO.
Symmetry under the mirror operation perpendicular to
the wave vector ${\bf q}$ forces
${\bf P}$ to be perpendicular to ${\bf q}$.
Although calculations at zero temperature[\onlinecite{ANGST}] indicate
that ${\bf P}$ is in the direction from a negative charge in one TLL
to a positive charge in the other TLL, this special result need not hold
for all temperatures and it would be interesting to actually measure
the direction of ${\bf P}$ as a function of temperature.  

In summary: in this Letter we have provided a phenomenological
explanation of the charge ordering, spin ordering, and
ferroelectricity in LuFe$_2$O$_4$.

We thank A. D.  Christianson and M. Angst for helpful correspondance,
A. Boothroyd for introducing us to this subject, and E. J. Mele
and T. C. Lubensky for several stimulating discussions.


\begin{thebibliography} {99}
\bibitem{FNT}
The temperatures of the phase transitions in LFO are
subject to some controversy.
\end{thebibliography}
\end{document}